\documentclass[10pt,conference]{IEEEtran}

\ifCLASSOPTIONcompsoc
  \usepackage[nocompress]{cite}
\else
  \usepackage{cite}
\fi

\usepackage{amsmath,amssymb,amsfonts}
\usepackage{algorithmic}
\usepackage{graphicx}
\usepackage{textcomp}
\usepackage{xcolor}
\def\BibTeX{{\rm B\kern-.05em{\sc i\kern-.025em b}\kern-.08em
    T\kern-.1667em\lower.7ex\hbox{E}\kern-.125emX}}
\usepackage{orcidlink}
\usepackage{pgfplots}
\usepackage{hyperref}
\usepackage{booktabs}
\usepackage{tabularx}
\usepackage{multirow}
\usepackage{tabularx}
\usepackage{xcolor}
\usepackage[pdf]{graphviz}
\usepackage{multicol, blindtext}
\usepackage{amsthm}
\usepackage{epigraph} 
\usepackage[leftmargin=0.1cm,rightmargin=0.1cm]{quoting}
\usepackage[inline]{enumitem}
\usepackage{pdfpages}

\newtheorem{example}{Example}

\usepackage[simplified]{pgf-umlcd}
\usepackage{tikzsymbols}
\usetikzlibrary{arrows}
\usetikzlibrary{positioning}
\usetikzlibrary{patterns}
\usetikzlibrary{arrows.meta}
\usetikzlibrary{
    calc, shadows,
    shapes,
    shapes.geometric,
    shapes.symbols,
    shapes.arrows,
    shapes.multipart,
    shapes.callouts,
    shapes.misc}
\usetikzlibrary{patterns}

\usepackage[first=-100, last=100]{lcg}

\usepackage{bp-listings}
\newcommand{\lstin}[2][]{\lstinline[style=,language=,basicstyle=\ttfamily,#1]|#2|}

\makeatletter
\def\lst@makecaption{%
  \def\@captype{table}%
  \@makecaption
}
\makeatother

\pgfplotsset{compat=1.18} 

\begin{document}

\title{Provengo: A Tool Suite for \\ Scenario Driven Model-Based Testing} 

\author{
Michael Bar-Sinai~\orcidlink{0000-0002-0153-8465},
Achiya~Elyasaf~\orcidlink{0000-0002-4009-5353},
Gera~Weiss~\orcidlink{0000-0002-5832-8768},
and Yeshayahu~Weiss~\orcidlink{0000-0002-8183-5282},
\IEEEcompsocitemizethanks{%
\IEEEcompsocthanksitem Y. Weiss and G. Weiss are with the Computer Science Department, Ben-Gurion University of the Negev, Israel.\protect\\
\IEEEcompsocthanksitem A. Elyasaf is with the Software and Information Systems Engineering Department, Ben-Gurion University of the Negev, Israel.\protect\\
E-mail: \texttt{achiya@bgu.ac.il}
\texttt{weissye@post.bgu.ac.il} and \texttt{geraw@cs.bgu.ac.il}}
}


\IEEEtitleabstractindextext{%

}

\maketitle


%

\begin{abstract}
We present Provengo, a comprehensive suite of tools designed to facilitate the implementation of Scenario-Driven Model-Based Testing (SDMBT), an innovative approach that utilizes scenarios to construct a model encompassing the user's perspective and the system's business value while also defining the desired outcomes. With the assistance of Provengo, testers gain the ability to effortlessly create natural user stories and seamlessly integrate them into a model capable of generating effective tests. The demonstration illustrates how SDMBT effectively addresses the bootstrapping challenge commonly encountered in model-based testing (MBT) by enabling incremental development, starting from simple models and gradually augmenting them with additional stories. \end{abstract}


\section{Introduction} \label{sec:intro}
Scenario-driven testing has proven to be highly effective in various tools for capturing the user perspective and ensuring a testing process that prioritizes user value~\cite{arnold2010scenario}. It enables realistic simulations of user interactions, leading to meaningful test cases that validate critical functionalities and user expectations. Its adoption attests to its success in enhancing the testing process and delivering software that meets user needs.


This paper introduces Provengo, a tool suite for Scenario-Driven Model-Based Testing (SDMBT). Setting it apart from other scenario-driven testing approaches, Provengo goes beyond using scenarios as standalone tests. Instead, it leverages scenarios as building blocks to construct a comprehensive model, enhancing the testing process. This innovative approach empowers testers to capture the user perspective and highlight the system's business value. Leveraging the power of model-based testing, Provengo provides scenario coverage, efficient testing through intelligent story interleaving, improved clarity of system behavior, and easy maintainability. With Provengo, testers can effortlessly create user stories that specify the base scenarios for testing the system. The tool then generates tests and produces insightful reports. Our demonstration showcases how Provengo effectively addresses testing challenges, including the bootstrapping issue in model-based testing~\cite{apfelbaum1997model}.

\section{An Overview of Provengo’s Modeling Approach}
\label{sec:poc}
Provengo, a test automation tool, is gaining a modest yet stable market share.
By providing the Provengo system with a few lines of instructions, QA users can automate and synthesize all interactions and communications related to product definition between all teams involved in the development process~\cite{provengo1}.

At the core of the Provengo tool is Behavioral Programming (BP) --- a modeling paradigm that enables developers to design complex reactive systems in an incremental and modular manner, where each module is aligned with a single aspect of the system behavior, preferably a requirement~\cite{Harel2010ProgrammingCoordinated, Harel2012BehavioralProgramming}.
Specifically, as its foundational technology, the Provengo tool harnesses the power of BPjs~\cite{BarSinai2018BPjs}, a framework for running and analyzing behavioral programs. By wrapping BPjs with test-oriented tools, Provengo empowers testers with a new model-based testing method.

Behavioral programming (BP) was chosen as the basis for the Provengo testing tool because it allows for direct alignment of the model with the project’s requirements~\cite{Elyasaf2020COBP}. The model is built of separate modules, each corresponding to an individual requirement. This is ideal for testing as it lets testers focus on individual requirements while the tool handles testing their combinations. For example, if a system has requirements for customer cart editing and store manager inventory editing, BP can help create tests that validate customer behavior even if inventory updates occur while they add products to their cart. This makes BP a powerful tool for ensuring that systems behave as expected under different conditions and scenarios.

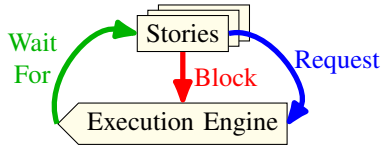
\begin{figure}
\centering
\begin{tikzpicture}[
doc/.style={draw, double copy shadow={shadow xshift=4pt,
shadow yshift=2pt, draw}},
, scale=0.6
]

\node [doc,fill=yellow!10!white] (scenarios) at (0,7) {Stories};
\node [signal,draw, signal to=west, fill=yellow!10!white] (engine) at (0,5) {Execution Engine};

\tikzset{>={Latex[round]}, line width=2pt}

\draw [->,green!70!black, line width=2pt] (engine.west) to [bend left=60] node [left, align=center, text width=.8cm] {Wait \\ For} (scenarios.west);
\draw [->,blue, line width=2pt] (scenarios.east) to [bend left=70] node [right, align=center] {Request} (engine.east);
\draw [->,red, line width=2pt] (scenarios) to node [right, align=center] {Block} (engine);
\end{tikzpicture}

\caption{A depiction of the execution cycle unfolds as follows: repeatedly, stories define events they wish to trigger and events they block. An application-agnostic execution engine selects an event that aligns with the requests and has not been blocked. Subsequently, the engine informs the stories that have been awaiting or requested for the selected event. The resulting test materializes as a sequence comprising the generated events in chronological order.}
\label{fig:execution-cycle}
\end{figure}

Provengo’s execution mechanism follows the standard behavioral programming run cycle, as shown in \autoref{fig:execution-cycle}. Stories are modeled using standard JavaScript procedures and are repeatedly run until they reach a synchronization point. This is marked by calling the \lstin{sync} method and declaring ``Request'', ``Wait For'', and ``Block'' events. The ``Execution Engine'' mechanism then selects a requested event that is not blocked, informs the requested or waited for the selected events, and continues running the stories. This cycle repeats until no more events can be triggered, defining a set of tests based on event selection at each round. 

To make the tool accessible to non-programmers, Provengo contains domain-specific modeling idioms implemented as an internal DSL (Domain-Specific Language) within JavaScript. An internal DSL is a type of programming language designed for a specific domain or task. It is implemented within a host language, in this case JavaScript, and allows users to express their intentions in a way that is more natural and intuitive for the specific domain.

\begin{example}
\footnotesize
Consider testing an application with two buttons, red and green. There are two stories: one requests to push the green button three times and the other requests to push the red button ten times (both modeled using a simple JavaScript loop). When run together, both stories progress to their first synchronization point where one requests to push the red button and the other requests to push the green button. Since there is no blocking, both events are possible and one is selected arbitrarily. This process repeats, resulting in a set of $1064$ possible runs, calculated by $\binom{13}{3}= 1064$, where $13$ is the total number of button presses ($3$ green plus $10$ red) and $3$ is the number of times the green button is pressed.
\end{example}

As this example demonstrates, stories can interleave to generate many tests. However, it is rare for all possible combinations to be allowed. The next example further illustrates this observation:

\begin{example}
\footnotesize
Provengo allows testers to use behavioral programming idioms to add new constraints to their model without modifying the original code. For example, a new rule can be added to prevent the green button from being pressed twice in a row by adding a new story that waits for the red button to be pressed before allowing the green button to be pressed again. This additionally decreases the number of possible test scenarios, in this case, to $(11 \text{ choose } 3)=165$ tests, as there are $11$ spaces between the ten presses of the red button to place the $3$ presses of the green button. 
\end{example}


The Provengo tool allows users to define high-level events that reflect different types of interactions with the system under test. These events can be used abstractly using a language that users and requirement engineers would normally use. For example, a user could define an event for adding an item to a cart and then use it in a scenario such as ``add items, checkout, and see that the price is OK.'' This approach makes it easier for users to create tests that reflect real-world scenarios and interactions with the system. By using a language familiar to users and requirement engineers, Provengo makes it easier for them to express their intentions and create tests that accurately reflect the behavior of the system under different conditions.

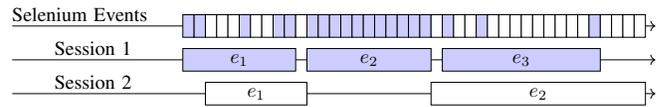
\begin{figure}

\scalebox{0.75}
{
\centering
\begin{tikzpicture}[
    every node/.style={inner sep=0,outer sep=0.7, anchor=west}
]
  \pgfmathsetseed{1984}

  \draw[->] (-3,3)   -- node[above=0.05cm,text width=3cm] {Selenium Events} (-0,3) -- (42*0.2,3);
  \draw[->] (-3,2.4) -- node[above=0.05cm,text width=1.5cm] {Session 1} (-0,2.4) -- (42*0.2,2.4);
  \draw[->] (-3,1.8) -- node[above=0.05cm,text width=1.5cm] {Session 2} (-0,1.8) -- (42*0.2,1.8);
  
  \foreach \x in {0, 1, ..., 40} {
    \pgfmathdeclarerandomlist{color}{{blue!20!white}{white}}
    \pgfmathrandomitem{\c}{color}

    \node[draw, minimum height=0.4cm, minimum width=0.2cm, fill=\c] at (\x*0.2, 3) {};
  }

  \node[draw, minimum height=0.4cm, minimum width=0.2*9cm, fill=white] at (0.2*2, 1.8) {$e_1$};
  \node[draw, minimum height=0.4cm, minimum width=0.2*19cm, fill=white] at (0.2*22, 1.8) {$e_2$};

  \node[draw, minimum height=0.4cm, minimum width=0.2*10cm, fill=blue!20!white] at (0, 2.4) {$e_1$};
  \node[draw, minimum height=0.4cm, minimum width=0.2*11cm, fill=blue!20!white] at (0.2*11, 2.4) {$e_2$};
  \node[draw, minimum height=0.4cm, minimum width=0.2*14cm, fill=blue!20!white] at (0.2*23, 2.4) {$e_3$};


\end{tikzpicture}
}
\caption{Execution semantics of low-level and high-level events in sessions: stories specify possible sequences of high-level events that run in parallel sessions (the bigger boxes in the last two timelines). Each high-level event is translated to a sequence of low-level events, and the interleaving is done at the level of low-level events (the first timeline). }
\label{fig:interleaving}
\end{figure}

The mechanism for integrating high- and low-level events is a process of interleaving, which allows for the coordination of both types of events in a parallel way as illustrated in \autoref{fig:interleaving}.
In this process, the stories specify possible sequences of high-level events that run in parallel sessions. These high-level events are represented as the bigger boxes in the last two timelines. Then, each high-level event is translated into a sequence of low-level events. This translation is necessary because high-level events are more abstract and less specific, while low-level events are more concrete and contain many details we want to encapsulate. The separation is necessary to make the tests more actionable, more understandable, and more resilient to changes. The interleaving is done at the level of low-level events, as seen in the first timeline. This means that the low-level events from each parallel session are coordinated so that they can run together without conflicts. This process allows for the efficient execution of high-level events and a more detailed view of low-level events.

It is worth noting that sessions in Provengo are not necessarily directly correlated with stories. One story can specify constraints for multiple sessions, such as blocking event X in session one until event Y in session two is triggered. Additionally, multiple stories can relate to the order of events in one session. For example, story one could specify triggering ten X events in a session, while story two specifies triggering three Y events in the same session. This flexibility allows developers to create rich test models by combining and reusing individual components and by specifying multiple constraints across different sessions. These ideas are elaborated and demonstrated in the following sections.

\begin{example}
\footnotesize
Provengo allows developers to modularly specify high-level events like ``HOT'' and ``COLD'' and their respective low-level events, e.g., ``$hot_1$'', $\dots$, ``$hot_3$'' and ``$cold_1$'', $\dots$ ``$cold_3$'', in a composable manner. These high-level events can be triggered in separate sessions to simulate real-world scenarios where multiple users or components may interact with the system simultaneously or with overlapping periods of activity. Provengo generates all possible combinations of these events, resulting in an exhaustive list of $70$ functional tests for the example system. This comprehensive test coverage ensures thorough testing of the system's behavior under a range of realistic conditions without the need for manual test case creation.
\end{example}

\begin{figure}
\centering

\includegraphics[width=0.5\textwidth]{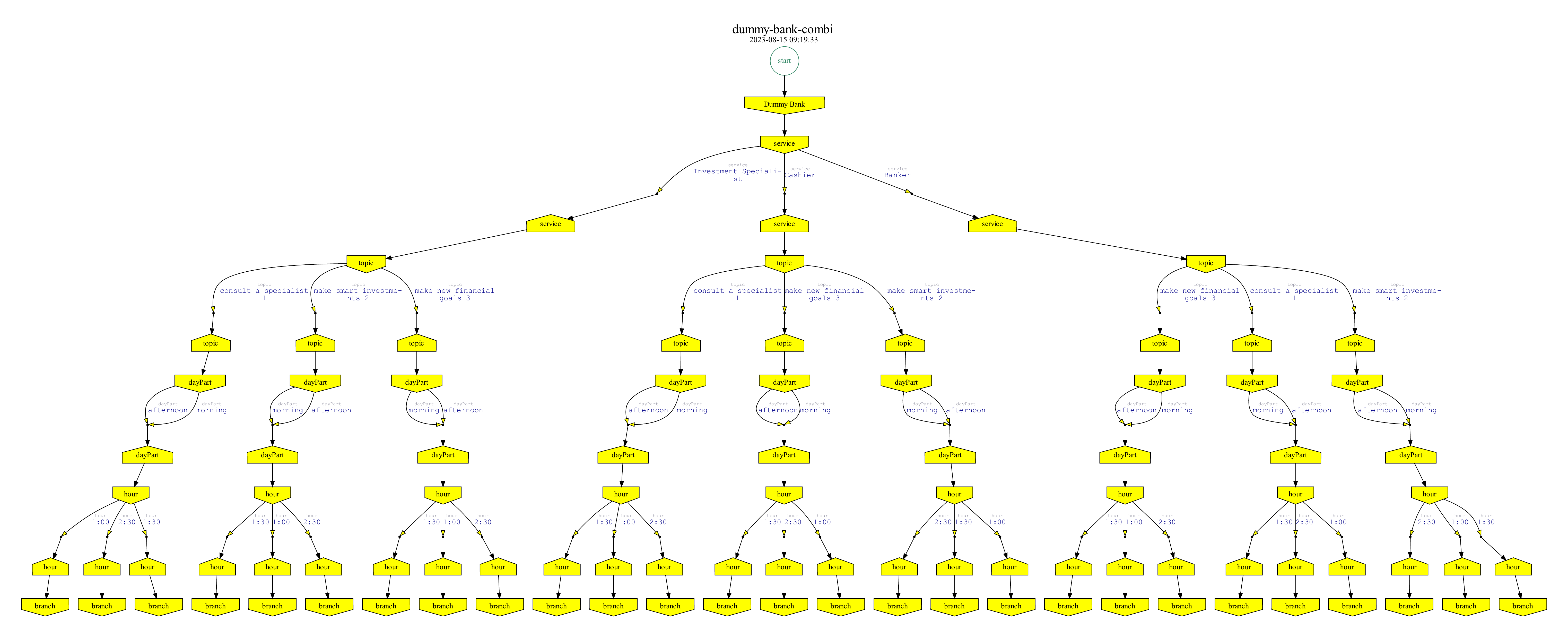}

\caption{A graphical abstraction of a test model (depth of 15) produced to allow feedback from users and requirement engineers.}
\label{fig:graph}
\end{figure}

\section{Leveraging Created Models: Diverse Approaches and Techniques}

Development teams with access to powerful model-based testing tools like Provengo can create comprehensive models that capture the verification steps required to ensure the system under test meets its requirements. These models generate a range of test scenarios, including intricate possibilities that testers may not have considered independently. 



One significant advantage of this approach is its capacity to enable testers to concentrate on test determination rather than being swamped by testing intricacies. Testers can define criteria such as requirement coverage metrics to steer their test selection, prioritizing tests likely to uncover system defects. Provengo furnishes the ability to produce model representations with adjustable levels of abstraction, demonstrated in \autoref{fig:graph}, as a crucial communication tool between testing teams and requirements engineers. This addresses a common challenge in software development, where the development team may not always prioritize user-aligned requirements. The tool facilitates early feedback on anticipated system behavior, effectively conveying test scenarios to requirements engineers and guiding the development team's focus.

Another advantage of leveraging models is the generation of targeted tests tailored to specific scenarios. For example, when replacing the database, it becomes crucial to concentrate on tests that cover tasks such as inventory backup and restoration. Similarly, tests focusing on currency exchanges become paramount when preparing to launch a product in a foreign country. Provengo supports this need by providing users with tools to specify factors they want to emphasize within their test suite. An optimization tool in Provengo extracts test suites from the model, maximizing the desired factors. This capability allows testers to prioritize tests for their specific requirements, thereby enhancing the overall effectiveness and efficiency of the testing process.

Furthermore, the models generated by Provengo facilitate effective communication of testing progress and the system's readiness for production to management. Informative reports can be generated, providing insights into the execution status of different test types. Statistical inference techniques can be applied to compute the probability of a random test failing within each test category, incorporating metrics such as variance and confidence levels. These comprehensive reports offer stakeholders valuable information regarding testing progress and the system's preparedness for release. These reports enable informed decision-making, enhance transparency, and facilitate effective communication between testing teams and management by presenting clear and quantifiable data.

Moreover, the testing models are powerful tools for localizing and understanding bugs' root causes. Through model analysis alongside corresponding test results, testers can precisely pinpoint the specific areas within the system where bugs occur and gain insights into potential contributing factors. This understanding significantly aids in diagnosing and resolving bugs efficiently, empowering testers to address the root causes and rectify underlying issues effectively. The bug resolution process becomes streamlined by leveraging the insights the testing models provide, resulting in improved software quality and enhanced system performance.

The tools provided by Provengo allow testers to harness the potential of created models and employ diverse approaches and techniques; software testers can elevate their testing efforts, improve test coverage, enhance bug detection, and ensure the system meets required quality standards. The Provengo model-based testing tools offer valuable capabilities that support effective communication, targeted testing, progress reporting, and bug diagnosis, ultimately contributing to the overall success of the software development life cycle.

\section{Quick-start Demonstration}
In this section, we present a demonstration illustrating the ease and potency of Provengo in testing Google's search engine. The process involves the creation of a simple story and an event that instructs Provengo on how to execute the test. Subsequently, we observe how Provengo generates and executes the test, providing a comprehensive result report.

\subsection*{Step 1: Event Definition}
The event definition is a directive to Provengo for performing specific actions on the desired website or application. These specific actions are a list of low-level events derived from a part of a story. The event is written in JavaScript, and each event definition corresponds to a step within the story. For instance, if a step in the story is labeled as ``ComposeQuery,'' an event definition of the same name is crafted to instruct Provengo to enter search text in the search box.

\begin{lstlisting}
defineEvent(SeleniumSession, "ComposeQuery", 
  function (session, event) {
    session.writeText("//input[@name='q']", event.text);
});

defineEvent(SeleniumSession, "StartSearch", 
  function (session, event) {
    session.click("//input[@name='btnK']");
});

defineEvent(SeleniumSession, "FeelLucky", 
  function (session, event) {
    session.click("//input[@name='btnI']");
});
\end{lstlisting}

The event definitions are constructed using Provengo's integration with Selenium\footnote{Selenium~\cite{Selenium} is an automated testing framework widely used for web application testing. Enabling developers to write test scripts that simulate user interactions with web browsers and validate expected functionalities.}, a popular web testing library. Selenium empowers interactions with web elements, such as buttons, links, and text boxes. In our illustration, we create two high-level event definitions: ``ComposeQuery'' and ``StartSearch,'' which use Selenium to type into the search box and initiate a search, respectively.

\subsection*{Step 2: Stories}


Stories act as a mechanism to communicate testing objectives to Provengo. These JavaScript-based narratives consist of steps that employ the previously defined events. Each step specifies the desired action or condition for Provengo to address. For instance, if the intention is to test Google's search engine, a suitable story may be ``Search for something on Google and check the results.''

Provengo has a robust capability of combining multiple stories to generate diverse test scenarios. Although this feature is not utilized in our present example, its functionality will be elucidated subsequently. For now, our focus remains on constructing a single story.

Currently, we craft a story called ``Search for Pizza.'' This story encompasses three events: ``StartSession,'' ``ComposeQuery,'' and ``StartSearch.'' The "StartSession" event is a built-in instruction for Provengo to initiate a new browser window and navigate to the designated website. The ``ComposeQuery'' and ``StartSearch'' events are those we defined earlier, dictating the input of search terms and the execution of the search.

\begin{lstlisting}
story('SearchPizzaOnGoogle', function () {
  with(new SeleniumSession().start('www.google.com')){
    composeQuery({ s: 'A1', text: 'Pizza' })
    startSearch({ s: 'A1' })
  }
});
\end{lstlisting}


A significant benefit of utilizing Provengo is the separation of testing logic from implementation specifics. The testing logic, embodied in the stories, articulates test objectives, while the event definitions encapsulate the implementation details governing web applications or site actions. This separation enables the creation of concise and lucid stories, free from the intricacies of technical web element interactions, which are adeptly managed by their corresponding event definitions.

\subsection*{Step 3: Using the model}
Provengo incorporates advanced tools rooted in recent developments in combinatorial sequence testing~\cite{elyasaf2023generalized}. These tools empower users to explore and test various scenarios based on the specified stories and events. A concise overview of these tools and their applications is provided below:

\paragraph*{Sampling} This feature facilitates randomly selecting a subset of scenarios from the project's specification space. Users can define the desired number of scenarios to sample, and Provengo generates a file containing these scenarios. The file can subsequently be utilized for test execution or highlighting within the model graph. The command for this feature is \lstin{provengo sample -n NUM <path-to-project>}, where \lstin{NUM} signifies the desired number of scenarios to be sampled.

\paragraph*{Ensembling} This functionality enables the creation of a test ensemble, comprising scenarios that cover diverse aspects of the project's specification space. Users can specify coverage, diversity, or complexity criteria to guide the ensemble's formation. Provengo then generates a file containing the scenarios belonging to the ensemble, which can be used for test execution or highlighting in the model graph. The command for utilizing this feature is \lstin{provengo ensemble -c CRITERIA <path-to-project>}, with \lstin{CRITERIA} indicating the chosen criteria for ensembling.

\paragraph*{Generating a Model Graph with Highlights} This tool visually represents the project's specification space as a graph. The graph depicts all possible scenarios stemming from the stories and events. Moreover, users can highlight a specific group of scenarios, such as those obtained through sampling or ensembling, to observe their relationship with the broader specification space. The graph's output format and style can be customized, with Provengo automatically launching it in the default viewer. To access this tool, the command is \lstin{provengo analyze [-f FORMAT] <path-to-project>}, where \lstin{FORMAT} designates the output format (json, gv, or pdf), and \lstin{PATH} corresponds to the path of a file containing the scenarios earmarked for highlighting.

\section{An Involved Example: Testing an Online Store}

We have also created a tutorial demonstrating how to utilize Provengo to test an online store. This  guide focuses on testing online store powered by the widely-used Magento e-commerce platform by Adobe. The tutorial covers various aspects of testing an online store, including interaction with the website using Selenium and the REST interface, and generating and executing test scenarios with Provengo using dynamic context data.
The tutorial consists of the following sections:
\begin{itemize}
    \item Event Definitions: Learn to define events using Selenium and REST for interacting with the website.
    \item Writing Stories: Discover how to write testable stories using Provengo's syntax and features, including dynamic elements like variables and conditions.
    \item Test Execution and Reporting: Execute tests using Provengo's engine and access detailed reports on the results through the command line interface.
    \item Advanced Features: Explore advanced features like sampling, ensembling, and model graph generation to enhance testing capabilities.
\end{itemize}
The tutorial can be found at \url{http://docs.provengo.tech/main/site/ProvengoCli/0.9.5}, and will be presented during the ASE 2023 tool demonstration. See a screencast of the demonstration at \url{https://provengo.tech/?page_id=196}. See also  \url{https://youtube.com/@provengo}. 

\section{A validation study}

 We conducted a case study to validate the effectiveness of the approach, focusing on testing the Magento e-commerce platform by extending the tutorial. As a result, two major bugs in the cart management logic were identified and reported to Adobe. The study highlights the effectiveness of scenario-driven model-based testing of a mature system, even in the absence of a formal specification, as it shows that testers can start with small models and gradually expand them based on their intuitive understanding of the system. The details of the case study are reported in a separate paper.


\bibliographystyle{IEEEtran}
\bibliography{bib}

\end{document}